\documentclass[10pt]{iopart}
\usepackage{iopams}  
\usepackage[xdvi]{graphicx}%

\newcommand{\bra}{\left\langle}
\newcommand{\ket}{\right\rangle}
\newcommand{\kets}[1]{\right\rangle^{{\rm s}}_{#1}}
\newcommand{\ketop}{\right\rangle^{\rm op1}}
\newcommand{\ketopp}{\right\rangle^{\rm op2}}
\newcommand{\Wop}{W^{\rm op1}}

\newcommand{\pder}[2]{\frac{\partial #1}{\partial  #2}}

\newcommand{\ps}{p_{\rm s}} 
\newcommand{\pc}{p_{\rm c}} 
\newcommand{\QR}{Q_{1}} 
\newcommand{\QRR}{Q_{2}} 
\newcommand{\Qr}{Q_{\rm b}}
\newcommand{\JR}{J_{\rm 1}} 
\newcommand{\Jr}{J_{\rm b}}

\newcommand{\Vt}{V_{\rm t}}
\newcommand{\Vint}{V_{\rm int}}
\newcommand{\Ftot}{F^{\rm mec}} 
\newcommand{\Veff}{V_{\rm eff}}
\newcommand{\Feff}{F^{\rm stat}}
\newcommand{\Fenergy}{F^{\rm thermo}}

\begin{document}

\title[Effective forces in nonequilibrium]
{The law of action and reaction for the effective force in a nonequilibrium 
colloidal system} 

\author{Kumiko Hayashi\dag\footnote[3]{To whom correspondence 
should be addressed (hayashi@jiro.c.u-tokto.ac.jp)} 
and Shin-ichi Sasa\ddag}

\address{\dag\ \ddag  
Department of Pure and Applied Sciences,  
University of Tokyo, Komaba, Tokyo 153-8902, Japan}

\begin{abstract}
We study a nonequilibrium Langevin many-body system containing two 
'test' particles and many 'background' particles.  The test particles are 
spatially confined by a harmonic potential, and the 
background particles 
are driven by an external driving force.  Employing 
numerical simulations of the model, we formulate  an effective 
description of the two test particles in a nonequilibrium 
steady state.  In particular, we investigate several different 
definitions of the effective force acting between the test particles.  
We find that 
the law of action and reaction does not hold for 
the  total mechanical force exerted by the background particles, but that 
it does hold for the thermodynamic force 
defined operationally on the basis of an idea used to extend the 
first law of  thermodynamics to nonequilibrium steady states.     
\end{abstract}



\maketitle

\section{Introduction}


  
Colloids are microscopic particles  
suspended in liquid.  Generally, they are on the order of   
micrometers in size, and thus they are much larger than 
molecules, but still small enough to exhibit Brownian motion.   
Colloids are convenient to study experimentally because 
unlike molecules, they can be easily observed with ordinary 
microscopes.  In recent years, as  the technology used in their 
manipulation has developed,  
experimental  studies of colloids have advanced greatly 
\cite{nature,bru1,bru2,colloid1,colloid2,HY}.  


In colloidal many-body systems, melting, freezing, glass transitions  
and nonequilibrium statistical mechanics have been studied.  
Among studies of  nonequilibrium steady states (NESSs),   
Dzubeilla et al.  investigated a model that describes  two fixed  
colloidal particles (test particles) in an environment containing  
many driven particles (background particles). They reported 
that  in a NESS, the effective interaction forces they defined 
between 
the two fixed particles violate the law of action and reaction 
\cite{DLL}.  This result implies that an effective  potential seems 
not to be constructed in NESS.  


The effective interaction force defined in Ref. [7] represents 
the two-body effect of the total force exerted by the background 
particles, which is extracted by subtracting from it the total 
force in the case that there is only colloidal particle. 
Although their definition seems plausible because of its simplicity, 
we expect that there is room  to define a different type of 
effective force in NESS.    
With this motivation,  we attempt to provide 
another idea for an effective interaction force between the two fixed
particles. 


In this paper, we reconsider effective forces in NESSs by studying a
system similar to that used in Ref \cite{DLL}.  First, we review 
the concept of effective forces in equilibrium  from 
three points of view,  statistical mechanics,  mechanics, and 
thermodynamics. Next, we investigate effective forces for nonequilibrium
systems by considering the results of  numerical experiments.  
Particularly, we 
demonstrate that in the NESS we consider, the law of action and 
reaction holds for 
the effective force defined according to a conjectured 
thermodynamic relation applied to NESSs.

\section{Model}\label{s:model}

\begin{figure}
\begin{center}
\includegraphics[width=8cm]{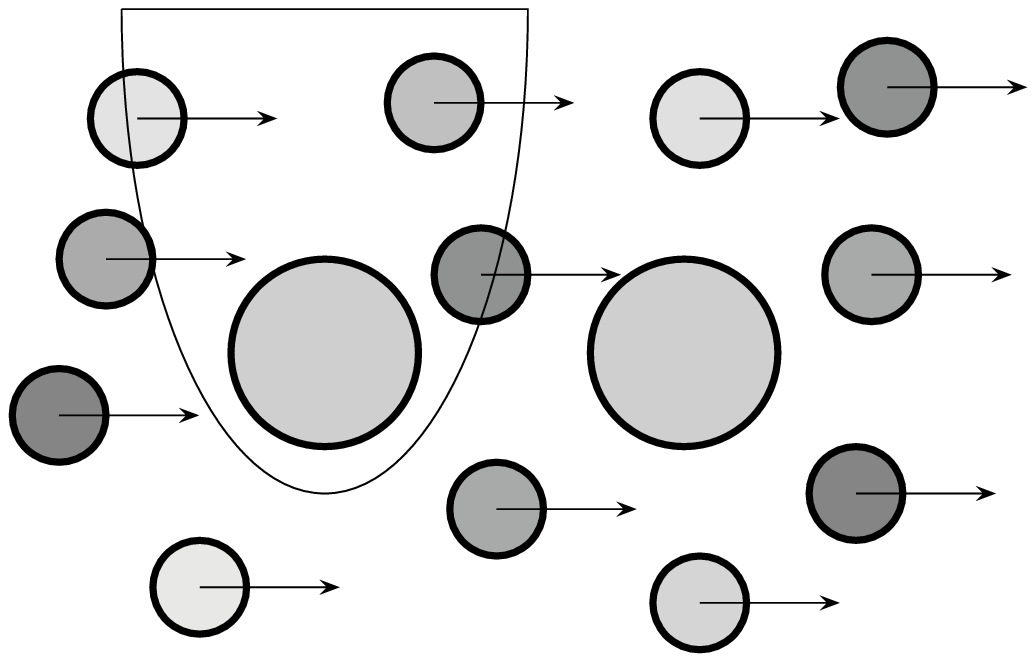}
\end{center}
\caption{Schematic representation of the model we study. 
Remark that  the background particles are not driven in the 
equilibrium case. 
}
\label{fig:colloid}
\end{figure}

The model we study describes a system  in which two 'test' particles
are trapped side by side in the center of the system 
by a harmonic potential and $N$ 'background' particles are driven by an 
external  force (see Fig. \ref{fig:colloid}).  
The test particles are unaffected by the driving force, and the 
background particles are unaffected by the trapping potential. 
Hereafter, we refer to the test particles  as particle {\bf 1} 
and  particle {\bf 2}. 

For the sake of simplicity, we consider the idealized situation of   
 a two-dimensional system  of length $L_x$ and $L_y$ in the $x$ 
and the $y$  directions,  with  periodic boundary conditions in both the 
directions. Then, letting $\vec R_1$ and $\vec R_2$ 
be the positions of particle {\bf 1}  and particle {\bf 2},  
we consider particle  
{\bf 2} fixed for simplicity, with  particle {\bf 1} trapped 
by the harmonic potential 
\begin{equation}
\Vt(\vec R_1)=\frac{k}{2}(\vec R_1-\vec R_{\rm c})^2, 
\end{equation}
where $\vec R_{\rm c}$ corresponds to the center of the potential, 
and $k$ is a spring constant chosen to be sufficiently large that 
particle {\bf 1} can be regarded as  almost fixed.  It is 
important to note 
that, although it may seem simpler to fix particle {\bf 1}, 
the dynamical degrees of freedom of particle {\bf 1} must be 
taken into account when we consider the distribution function of 
the effective description in which the degrees of freedom of the 
background particles are integrated. [If particle {\bf 1} is also 
fixed, no variable remains in the distribution function of the 
effective description (see the next section)].  
As a further simplification, it is assumed that there is no 
direct interaction between  particles {\bf 1} and  {\bf 2} and 
that there is no direct interaction among the  background particles.  
Then, we stipulate that the interaction  between particle 
{\bf 1} or particle {\bf 2} and  a background particle is 
represented by a short-ranged potential function 
$\Vint(\vec R_i-\vec r_k)$,    
where $\vec r_k$ is the position of the $k$-th background particle 
($k=1,\cdots,N$),  of the form 
\begin{equation}
\Vint(\vec r)=V_0\exp\left[-2\left|\frac{r}{\sigma}\right|^2\right].  
\label{pote5}
\end{equation}

With the system as described above, 
the motion of particle {\bf 1} is described by the Langevin equation 
\begin{equation}
\gamma \dot{\vec R_1}= -k(\vec R_1-\vec R_{\rm c}) 
-\sum_{k=1}^{N}\frac{\partial \Vint(\vec R_1-\vec r_k)}{\partial\vec R_1}
+\vec \xi(t), 
\label{model1}
\end{equation}
where $\gamma$ is a friction constant, and 
$\vec \xi(t)=(\xi_{x}(t), \xi_{y}(t))$ represents Gaussian white noise that  
satisfies   
\begin{equation}
\bra \xi_{\alpha}(t)\xi_{\alpha'}(t') \ket 
= 2 \gamma T \delta(t-t')\delta_{\alpha,\alpha'}.  
\label{noise}
\end{equation}
Here, the Boltzmann constant is set to unity,  and $T$ $(=1/\beta)$ is 
the temperature of the environment.  The motion of the $k$-th 
background particle is described by 
\begin{equation}
\gamma \dot{\vec r_k}=\vec f  
-\sum_{i=1}^2\frac{\partial \Vint(\vec R_i-\vec r_k)}{\partial\vec r_k}
+\vec \xi_k(t), 
 \label{model2}
\end{equation} 
where $\vec f=(f,0)$ is an external driving force, and   
$\vec \xi_k(t)=(\xi_{k,x}(t), \xi_{k,y}(t))$ is  Gaussian white 
noise that satisfies   
\begin{equation}
\bra \xi_{k,\alpha}(t)\xi_{k',\alpha'}(t') \ket 
= 2 \gamma T \delta_{k,k'}\delta(t-t')\delta_{\alpha,\alpha'}.  
\label{noise2}
\end{equation}

Below, all the quantities are converted into dimensionless forms 
by normalizing $\sigma$, $\gamma$ and $T$ to unity.  We set the parameters 
used for our numerical experiments as $L_x=11$, $L_y=6$, $N=200$, 
$k=50$, $V_0=7$, $0\le f\le 0.3$,  $\vec R_{\rm c}=(3,3)$ and  
$\vec R_2=(4.5,3)$. Also, all the numerical results in this paper 
is obtained by a finite difference method of the Langevin 
equations with a time step $2.5\times 10^{-4}$.

\section{Equilibrium case}

\subsection{Review of the effective description}

Before analyzing nonequilibrium cases, 
we review the {\it effective description} of 
particles  {\bf 1} and {\bf 2} in equilibrium,  
in which the stationary distribution function of the model 
is the canonical distribution 
\begin{equation}
p_{\rm c}(\vec R_1,\{\vec r_k\};\vec R_2)
=\frac{\exp\{-\beta[\Vt(\vec R_1)+\sum_{i=1}^2 \sum_{k=1}^N
\Vint(\vec R_i-\vec r_k)]\} }{Z}, 
\label{eqdis}
\end{equation}
where $Z$ is a normalization constant. 
In the equilibrium case, employing the well-established framework of 
statistical 
mechanics, we can derive the effective Hamiltonian of 
particles  {\bf 1} and {\bf 2}  by integrating (\ref{eqdis})  over the 
degrees of freedom of the background particles.

For the canonical distribution 
$p_{\rm c}(\vec R_1,\{\vec r_k\};\vec R_2)$ given by (\ref{eqdis}) 
and the interaction given by (\ref{pote5}), there exists a function 
$\vec R_1-\vec R_2$, which we write  $\Veff(\vec R_1-\vec R_2)$, 
satisfying the equation 
\begin{equation}
\exp\{-\beta[\Veff(\vec R_1-\vec R_2)+\Vt(\vec R_1)]\} =
\int \prod_{k=1}^{N} d\vec r_k p_{\rm c}(\vec R_1,\{\vec r_k\};\vec R_2). 
\label{effpote}
\end{equation}
It is natural to refer to the function $\Veff(\vec R_1-\vec R_2)$ 
as the effective interaction potential.  The fact that this effective 
interaction potential depends on only the relative displacement 
$\vec R_1-\vec R_2$  is due to the the spatial symmetry of the system. 
Then, as the left-hand side of (\ref{effpote}) depends on only 
$\vec R_1$ and $\vec R_2$, it is interpreted as the steady state 
distribution for the effective description: 
\begin{equation}
P_{\rm eff}(\vec R_1;\vec R_2)
=\exp\{-\beta[\Vt(\vec R_1)+\Veff(\vec R_1-\vec R_2)]\}.  
\label{ecan}
\end{equation}
Clearly the effective potential $\Veff(\vec R_1-\vec R_2)$  obtained 
in the above manner is properly defined, as this 
$P_{\rm eff}(\vec R_1;\vec R_2)$ is in canonical form. 

With the above argument establishing the proper definition of 
 $\Veff(\vec R_1-\vec R_2)$, 
 the {\it statistical mechanical force}  
between particles  {\bf 1} and  {\bf 2} is defined naturally by 
\begin{equation}
\vec \Feff_{i} \equiv  
-\frac{\partial \Veff(\vec R_1-\vec R_2)}{\partial \vec R_i},
\label{efffor}
\end{equation}
with $i=1,2$. Because $\Veff(\vec R_1-\vec R_2)$ is a function of 
the relative displacement  $\vec R_1-\vec R_2$,  
we immediately find the relation 
\begin{equation}
\vec \Feff_{1}=-\vec \Feff_{2},
\label{ar}
\end{equation}
which represents  the law of action and reaction for 
statistical mechanical 
forces in equilibrium. The statistical mechanical force defined above 
is closely 
related to the {\it mechanical force} exerted by all the background particles, 
\begin{equation}
\vec \Ftot_{i} \equiv   \sum_{k=1}^N 
\frac{\partial \Vint(\vec R_i-\vec r _k)}{\partial \vec r_k}.
\label{efffor2}
\end{equation}
Indeed, using (\ref{effpote}) and (\ref{efffor}), we can prove the relation
\begin{equation}
\bra \vec \Feff_{i} \kets{f=0}= \bra \vec \Ftot_{i} \kets{f=0},
\label{relation1}
\end{equation}
where $\bra \ \kets{f=0}$ represents the statistical average
in the steady state with $f=0$ (see Subsection 3.1 for 
the proof of (\ref{relation1})).  From (\ref{ar}) and (\ref{relation1}),
we obtain the law of action and reaction for $\vec\Ftot_{i}$:
\begin{equation}
\bra \vec \Ftot_{1} \kets{f=0}=-\bra \vec \Ftot_{2}\kets{f=0}.
\label{ar:tot}
\end{equation}

\begin{figure}
\begin{center}
\includegraphics[width=8cm]{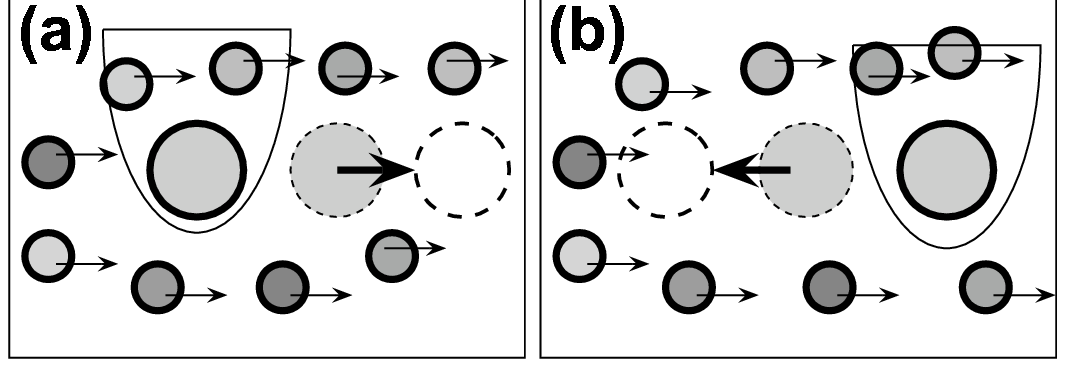}
\end{center}
\caption{(a) Simple operation shifting  
the position of particle {\bf 2} as  $\vec R_2 \to \vec R_2 
+  \Delta \vec R_2$ instantaneously at time $t=0$. 
Remark that  the background particles are not driven 
in the equilibrium case.
\\ 
(b) Simple operation in which  the roles of particles {\bf 1} 
and {\bf 2} in (a) are exchanged.  
}  
\label{fig:colloid3}
\end{figure}

The statistical mechanical force 
in equilibrium can also be understood 
within a  thermodynamic framework by considering 
 the case in which an operation 
is performed  on the system. We now  explain this understanding. 
As an example,  we consider the simple operation of shifting   
the position of particle {\bf 2} as   
$\vec R_2 \to \vec R_2 +  \Delta \vec R_2$ instantaneously
at time $t=0$.   (See Fig. \ref{fig:colloid3} (a) depicting the case $f=0$.)  
After the application of 
this operation,  the system is left  until a new equilibrium state 
is realized. Then, the work done by the external agent carrying out 
this operation, $\Wop$,  
satisfies the first law of thermodynamics: 
\begin{eqnarray}
&& \Wop =
\bra \QR\ketop_{f=0}+\bra \Qr\ketop_{f=0}
+\bra\Delta\Vt(\vec R_1)\ketop_{f=0}  \nonumber \\ 
&+&\bra\sum_{i=1}^2\sum_{k=1}^N \Delta \Vint(\vec R_i- \vec r_k)\ketop_{f=0}.
\label{energyb2}
\end{eqnarray}
Here $\bra \ \ketop_f$ represents the statistical average with respect to  
this operation.  The quantities $\QR$ and $\Qr$ are the 
energies transfered to 
the heat bath from the degrees of freedom of particle {\bf 1}
and the background particles. We refer to each of  these transferred 
quantities of energy as {\it heat}.  According to 
Ref.  \cite{sekimoto}, the quantities 
\begin{eqnarray}
\JR &\equiv& \left( \gamma  \dot{\vec R_1}-\vec \xi \right)
\circ \dot{\vec R_1} ,
\label{heat1} \\
\Jr&\equiv& \sum_{k=1}^N 
\left(\gamma \dot{\vec r_k}-\vec \xi_{k} \right)\circ 
\dot{\vec r_k}.
\label{heat2}
\end{eqnarray} 
represent the heat  transferred per the unit time, and therefore 
$\QR$ and $\Qr$ can be defined  as 
\begin{eqnarray}
\QR &\equiv& \int_0^{\infty} dt \JR(t), \\
\Qr &\equiv& \int_0^{\infty} dt \Jr(t).
\end{eqnarray}
The symbol $\circ$ in (\ref{heat1}) and  (\ref{heat2})  
indicates that the product here is the stochastic 
Stieltjes integral in the Stratonovich sense \cite{sekimoto}. 

Now, using the work, $\Wop$,  we define 
the {\it thermodynamic force}  by 
\begin{equation}
\vec \Fenergy_2 \equiv \lim_{|\Delta \vec R_2|\to 0}
\left(-\frac{\Wop }{\Delta (\vec R_2)_x}, -\frac{\Wop }{\Delta (\vec R_2)_y}
\right).
\label{Fenedef}
\end{equation}
Then, integrating the Langevin equations  
(\ref{model1}) and (\ref{model2}), and using the expression  
\begin{equation}
\dot{\vec R_2}=\Delta \vec R_2\delta(t)    
\label{ope}
\end{equation} 
to represent the operation, we can prove the relation 
\begin{equation}
\vec \Fenergy_2 = \bra\vec \Ftot_2\kets{f=0}, 
\label{fene2}
\end{equation}
(see Subsection 3.2 for the proof of (\ref{fene2})).

In the same manner that we 
 defined the thermodynamic force  $\vec \Fenergy_2$,
we can also define $\vec \Fenergy_1$ by exchanging  the roles of  
particles {\bf 1} and {\bf 2}.  (See Fig. \ref{fig:colloid3} (b) 
depicting  the  case $f=0$.)  
That is,   particle {\bf 1} is fixed, while particle {\bf 2} is 
trapped by the harmonic potential.  Because this exchange is 
equivalent to the spatial reflection with respect to $\vec R_2/2$,  
we find 
\begin{equation}
\vec \Fenergy_{1}=-\vec \Fenergy_{2}. 
\label{ar:ene}
\end{equation}
Then, from (\ref{ar:tot}), (\ref{fene2}) and (\ref{ar:ene}), we derive  
\begin{equation}
\vec \Fenergy_1 =\bra \vec \Ftot_1\kets{f=0}. 
\label{fene1}
\end{equation}

From the above results, we conclude that  
 in the equilibrium case,  because $\bra\vec \Feff_i\kets{f=0} 
=\bra\vec \Ftot_i\kets{f=0}=\vec\Fenergy_i$, 
we can regard  any one of these quantities 
as the {\it effective interaction force}  
between particles {\bf 1} and  {\bf 2}.

\subsection{Proofs of (\ref{relation1}) and (\ref{fene2})}

\subsubsection*{Proof of (\ref{relation1}):} 

The differentiation of (\ref{effpote}) with respect to $\vec R_2$ yields
\begin{eqnarray} 
&&\frac{\partial \Veff(\vec R_1-\vec R_2)}{\partial \vec R_2}
\exp\{-\beta[\Veff(\vec R_1-\vec R_2)+\Vt(\vec R_1)]\}\nonumber\\
&=&
\int\prod_{k=1}^{N}d\vec{r_k}\sum_{k=1}^{N} 
\frac{\partial \Vint(\vec R_2 -\vec r_k)}{\partial \vec R_2}
\pc(\vec R_1,\vec r_k;\vec R_2).
\label{ap1:1}
\end{eqnarray} 
Integrating this with respect to $\vec R_1$,  we derive 
\begin{eqnarray}
\bra \frac{\partial \Veff(\vec R_1 -\vec R_2)}{\partial \vec R_2}\kets{f=0}
&=& 
\bra \sum_{k=1}^{N} 
\frac{\partial \Vint(\vec R_2-\vec r_k)}{\partial \vec R_2 }\kets{f=0} \\
&=& 
-\bra \sum_{k=1}^{N} 
\frac{\partial \Vint(\vec R_2-\vec r_k)}{\partial \vec r_k }\kets{f=0}.
\label{ap1:2}
\end{eqnarray}
Using the definitions (\ref{efffor}) and 
(\ref{efffor2}), we obtain (\ref {relation1}).

\subsubsection*{Proof of (\ref{fene2}):} 
  
From  (\ref{model1}) 
and (\ref{model2}) with $f=0$, we obtain the energy balance equations:
\begin{eqnarray}
&&(\gamma \dot{\vec R_1}-\vec \xi)\circ \dot{\vec R_1}
= -\frac{\partial \Vt(\vec R_1)}
{\partial\vec R_1}\circ \dot{\vec R_1}
-\sum_{k=1}^{N}\frac{\partial \Vint(\vec R_1-\vec r_k)}{\partial\vec R_1}
\circ \dot{\vec R_1}
\label{ap2:1}
\\
&&(\gamma \dot{\vec r_k}-\vec \xi_k(t))\circ \dot{\vec r_k}
=-\sum_{i=1}^2\frac{\partial \Vint(\vec R_i-\vec r_k)}{\partial\vec r_k}
\circ \dot{\vec r_k}, 
\label{ap2:2}
\end{eqnarray}
where $k=1,\cdots,N$. The addition of (\ref{ap2:1}) and (\ref{ap2:2}) with 
$k=1,\cdots,N$ leads to 
\begin{eqnarray}
J_1+J_{\rm b}&=&-\sum_{i=1}^{2}\sum_{k=1}^N \dot{V}_{\rm int}(\vec R_i-\vec r_k)
-\dot{V}_{\rm t}(\vec R_1)\nonumber\\
&+& \sum_{k=1}^N \frac{\partial \Vint(\vec R_2-\vec r_k)}{\partial\vec R_2}
\cdot\Delta \vec R_2\delta(t),
\label{ap2:3}
\end{eqnarray}
where we have used (\ref{ope}). 
Here, we perform the time integration and consider its the statistical 
average. Noting that the third term of the right hand side of 
(\ref{ap2:3}) becomes $-\bra \vec \Ftot_2\kets{f=0} \cdot \vec \Delta R_2$,
we obtain
\begin{eqnarray}
&&-\bra \vec\Ftot_2\kets{f=0}\cdot\Delta \vec R_2
=\bra Q_1\ketop_{f=0}+\bra Q_{\rm b}\ketop_{f=0}\nonumber \\
&+&\bra \sum_{i=1}^{2}\sum_{k=1}^N \Delta{\Vint}(\vec R_i-\vec r_k)
\ketop_{f=0}+ \bra\Delta\Vt(\vec R_1)\ketop_{f=0}. 
\label{ap2:4}
\end{eqnarray}
Comparing this with the definition of 
$\Fenergy_2$ given in (\ref{energyb2}), we obtain 
\begin{equation}
\vec\Fenergy_2=\bra\vec\Ftot_2\kets{f=0}. 
\end{equation}

\section{Nonequilibrium case}

In this section, which consists of three subsections, 
 we study the nonequilibrium case.  We first note  
that the canonical distribution cannot be employed in the derivation 
of an effective description for a NESS, in contrast to the equilibrium 
case.  In the case of NESSs, 
 the relations among $\Feff_i$, $\Ftot_i$ 
and  $\Fenergy_i$ are not yet understood.  In this section, 
we investigate these three forces as defined above in the  
nonequilibrium, $f\ne 0$ case. We hope that this will allow us to 
obtain the proper effective description of particles {\bf 1} and 
{\bf 2} in a NESS.  
We first present the numerically computed results determining 
$\Ftot_i$ and $\Fenergy_i$ as defined in (\ref{efffor2}) and 
(\ref{Fenedef})   
for a  NESS.   Then, we derive $\Feff_i$ from the 
expression of the steady state distribution  
and compare  this $\Feff_i$ with $\Ftot_i$ and  $\Fenergy_i$. 
  
Throughout this section, we consider only the case of forces in the $x$ 
direction  
and  denote  the $x$ component of $\vec \Ftot_{i}$,  
$\vec \Fenergy_{i}$ and $\vec \Feff_{i}$ 
as $\Ftot_{i}$, $\Fenergy_{i}$ and $\Feff_{i}$.

\subsection{Mechanical forces exerted by the background particles}
\label{s:sum}

%
%
\begin{figure}
\begin{center}
\vspace{7mm}
\includegraphics[width=8cm]{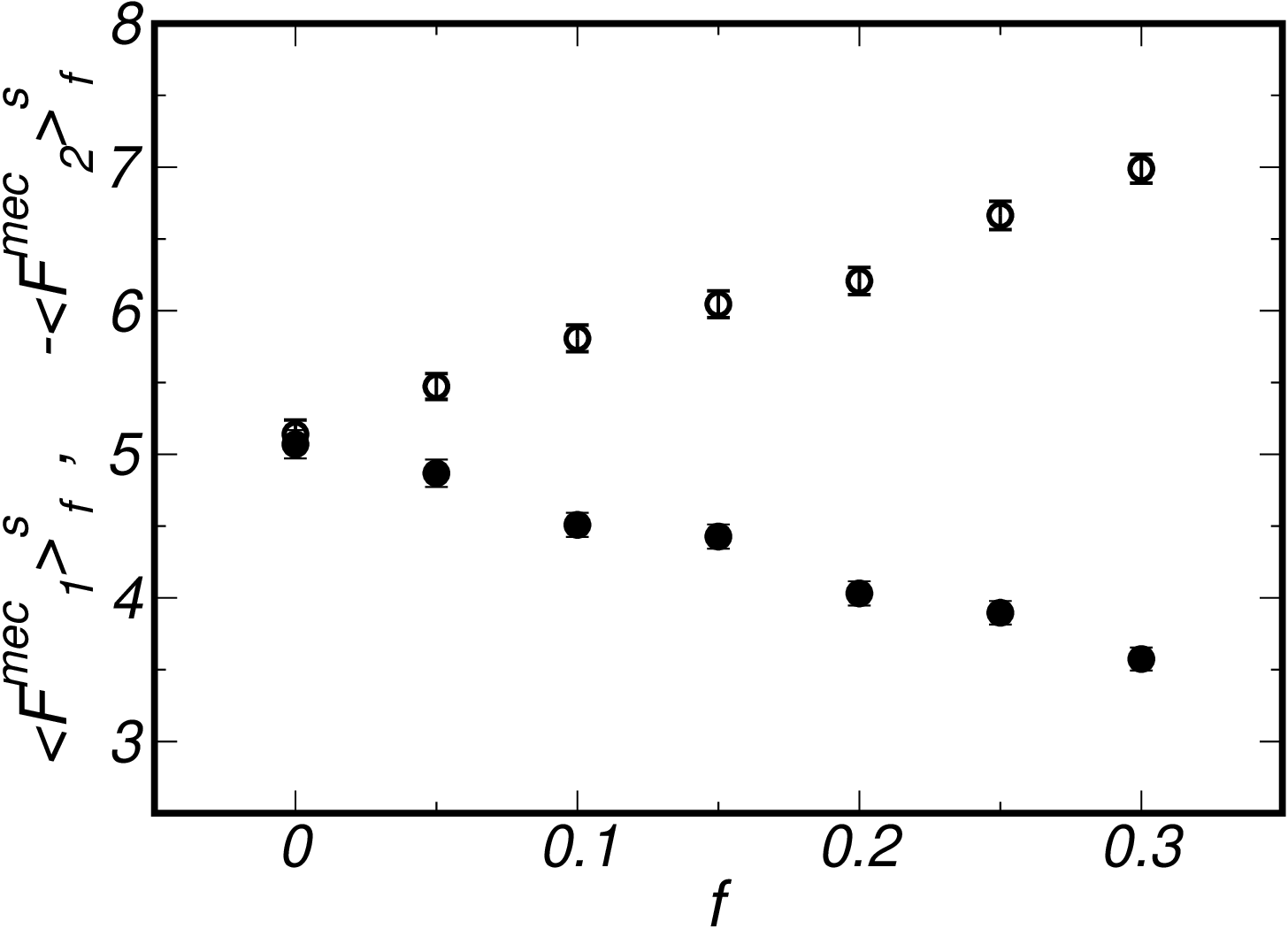}
\end{center}
\caption{$\langle\Ftot_1\rangle^{\rm s}_{f}$ 
(open symbols) and $-\langle\Ftot_2\rangle^{\rm s}_{f}$ 
(solid symbols) are plotted as functions of $f$. 
}
\label{fig:sum}
\end{figure}

In Fig. \ref{fig:sum}, $\bra\Ftot_1\kets{f}$ and $-\bra\Ftot_2\kets{f}$
are plotted as functions of $f$. In the case $f=0$, it is seen that
$\bra\Ftot_1\kets{f=0}=-\bra\Ftot_2\kets{f=0}$, which is consistent
with (\ref{ar:tot}).  However,  it is seen that for $f\ne 0$, these 
quantities are not equal, and indeed the discrepancy between them  
increases monotonically as a function of $f$. Hence, the law of 
action and relation does not hold with respect to $\bra\Ftot_1\kets{f}$ 
and $\bra\Ftot_2\kets{f}$ in the nonequilibrium case, unlike in the 
equilibrium case. 
 This asymmetry is interpreted 
as follows.  Particle {\bf 2} is partially shielded from the effect of 
the background particles by particle {\bf 1} in the case $f>0$ 
(see Fig. \ref{fig:colloid}).

\subsection{Thermodynamic definition}

Let us  now reconsider  the first law of thermodynamics represented by  
(\ref{energyb2}), which provides the definition of $\vec\Fenergy_{i}$ 
($i=1,2$).  In contrast to equilibrium cases, (\ref{energyb2}) 
cannot be employed directly 
 for NESSs, because $\Jr$,   
defined by (\ref{heat2}), takes a nonzero value even when there 
is no operation applied to the system, and hence $\Qr$ becomes infinite.   
(Recall that the background particles are driven by the external 
driving force, $f$. ) 

In order to extend  thermodynamic considerations to NESSs, we introduce
the idea of the {\it net heat} generated by an operation, following the  
framework of {\it steady state thermodynamics} proposed by 
Oono and Paniconi  \cite{oono}.   
In this framework,  the heat necessary to maintain 
a steady state (the so-called {\it house keeping heat}) is regarded as 
being independent of the change 
undergone by the system 
caused by the operation, and the 
{\it excess heat}  is defined by subtracting the house-keeping 
heat from the total heat dissipated to the heat bath.  
They conjecture that thermodynamics can be extended  
to  NESSs by using  this excess heat.   

\begin{figure}
\begin{center}
\vspace{7mm}
\includegraphics[width=8cm]{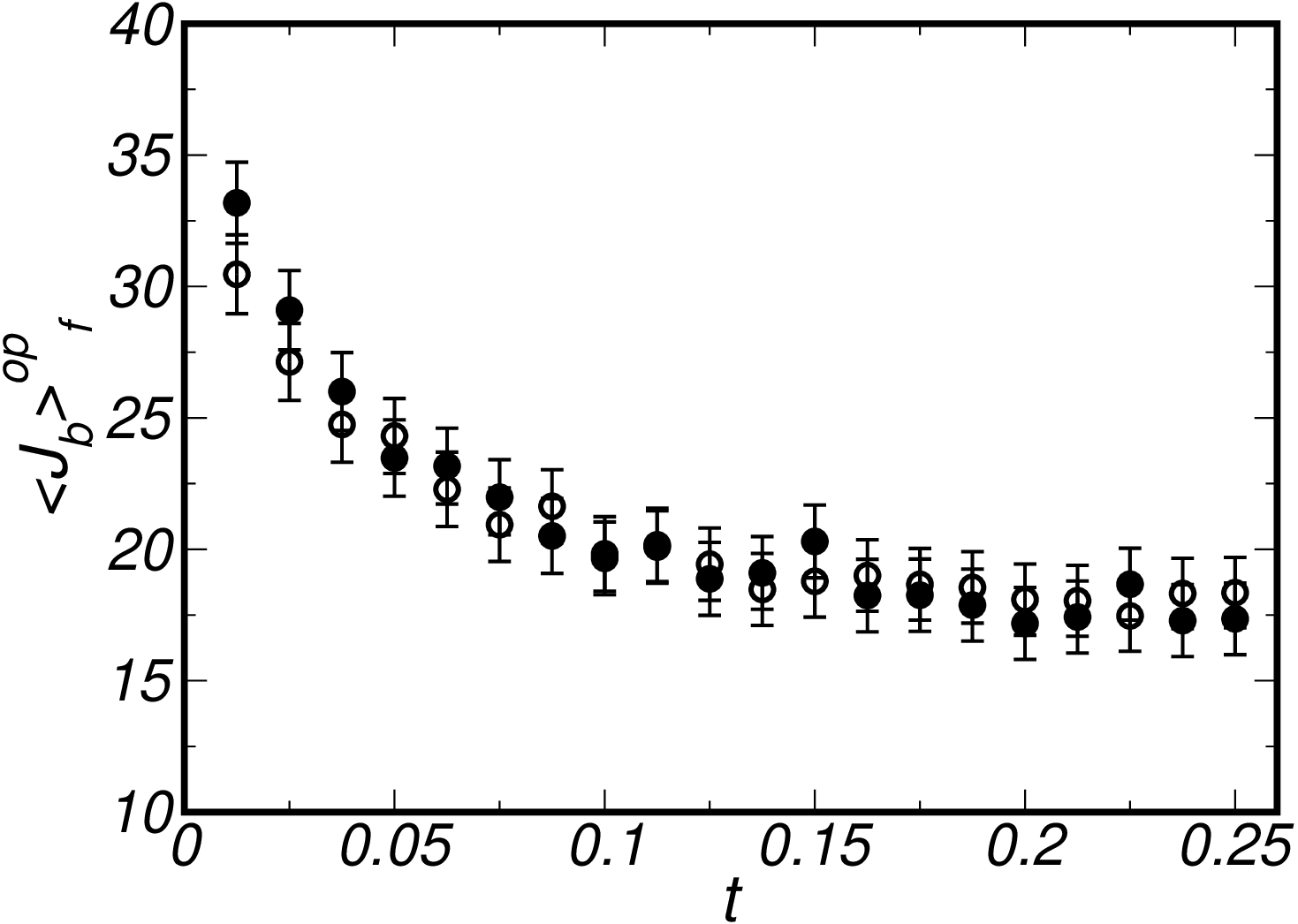}
\end{center}
\caption{$\bra\Jr(t)\ketop_f$  (open symbols)  
and  $\bra\Jr(t)\ketopp_f$ (solid symbols)   
 as functions of time $t$ 
in the nonequilibrium case with $f=0.3$.  Here, $\Delta R_2=0.25$ and 
$\Delta R_1=0.25$.  $\bra\Jr(t)\ketop_f$ and $\bra\Jr(t)\ketopp_f$ are  
computed in the situations described in Figs. \ref{fig:colloid3} (a) and   
 (b),  respectively.   
}  
\label{fig:heat0.3}
\end{figure}

Here, as one example, in Fig. \ref{fig:heat0.3}, 
$\bra \Jr(t) \ketop_f$ is plotted as a function of time in the case 
that $\Delta \vec R_2=(\Delta R_2,0)=(0.25,0)$ and  $f=0.3$.  
This graph shows that $\bra \Jr(t) \ketop_f$ 
reaches a steady state value, $\Jr(\infty)$,  sufficiently 
long time after 
the application of the 
operation. Then, employing the idea of the excess heat,  
it is reasonable to define  
\begin{equation}   
\tilde{Q}_{\rm b} \equiv  \int_0^\infty dt(\Jr(t) - 
\Jr(\infty)).   
\label{heat3}
\end{equation}
By replacing $\Qr$   in (\ref{energyb2}) 
with $\tilde{Q}_{\rm b}$,   
we hypothesize the following 
an extended form of the first law of thermodynamics: 
\begin{eqnarray}
&& \Wop =
\bra \QR\ketop_f+\bra \tilde{Q}_{\rm b}\ketop_{f}  
+\bra\Delta\Vt(\vec R_1)\ketop_{f} \nonumber \\  
&+&\bra\sum_{i=1}^2\sum_{k=1}^N \Delta \Vint(\vec R_i- \vec r_k)\ketop_{f}.
\label{energyb3}
\end{eqnarray}
Using this $\Wop$,  we define the {\it thermodynamic force} in 
the present NESS by (\ref{Fenedef}). 

\begin{figure}
\begin{center}
\vspace{7mm}
\includegraphics[width=8cm]{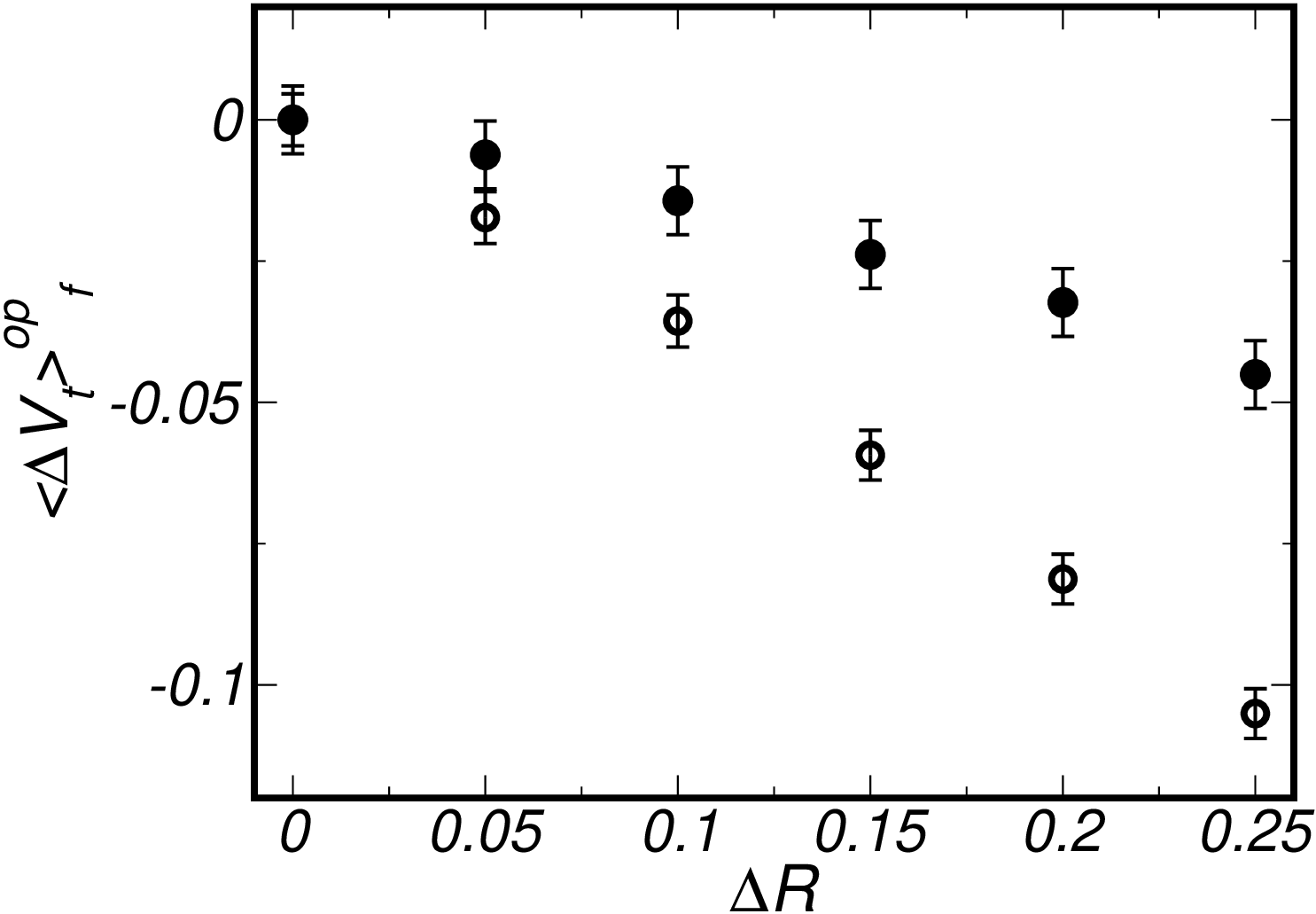}
\end{center}
\caption{$\bra\Delta \Vt\ketop_f$ as a function of $\Delta R_2$, and   
$\bra\Delta\Vt\ketopp_f$ as a function of $\Delta R_1$.  
  The open symbols and the solid symbols represent 
the quantities obtained in the situations described  
in Figs. \ref{fig:colloid3} (a)  and  (b),  
respectively.  
}  
\label{fig:ut}
\end{figure}

\begin{figure}
\begin{center}
\vspace{7mm}
\includegraphics[width=8cm]{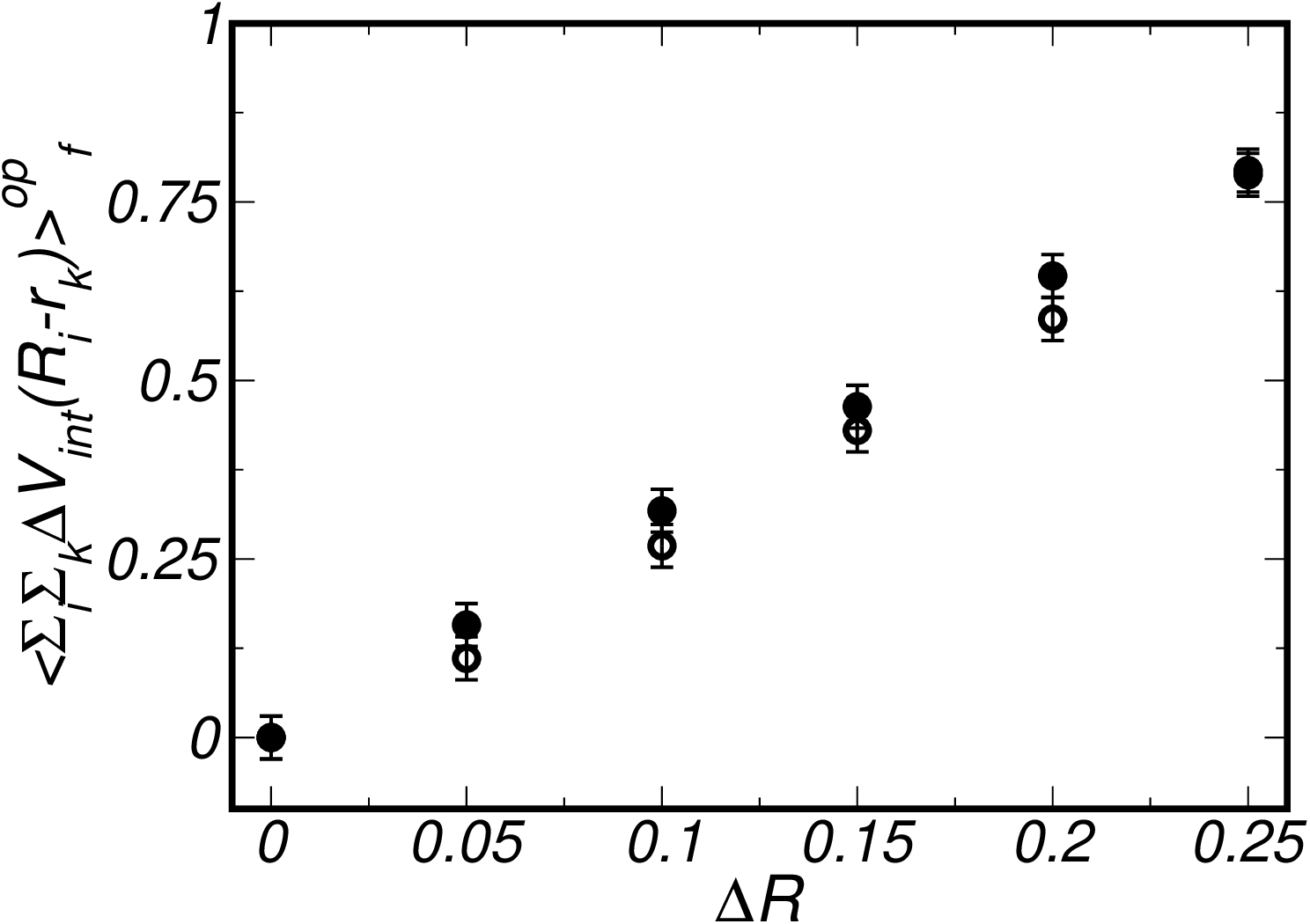}
\end{center}
\caption{$\bra\sum_{i=1}^2\sum_{k=1}^N\Delta\Vint(\vec{R}_i-\vec{r}_k)
\ketop_f$  as a function of $\Delta R_2$, and  
$\bra\sum_{i=1}^2\sum_{k=1}^N\Delta\Vint(\vec{R}_i-\vec{r}_k)
\ketopp_f$   as a function of $\Delta R_1$.  
 The open symbols and the solid symbols represent   
the quantities obtained in the situations described  
in Figs. \ref{fig:colloid3} (a)  
and  (b), respectively. 
}
\label{fig:u1u2}
\end{figure}

In Figs.  \ref{fig:ut} and \ref{fig:u1u2}, we plot 
$\bra\Delta\Vt\ketop_f$  and 
$\bra\sum_{i=1}^2\sum_{k=1}^N \Delta \Vint(\vec R_i- \vec r_k)\ketop_f$  
as  functions of $\Delta R_2$ in the case $f=0.3$. 
It is seen that these quantities depend linearly  on 
 $\Delta R_2$ in the range plotted. This implies that $\Delta R_2=0.25$ 
is  sufficiently small to be used  in (\ref{Fenedef}) to numerically 
compute the force. Using this value, we obtain  
$\bra \tilde{Q}_{\rm b}\ketop_f=0.61$, 
  $\bra\Delta\Vt(\vec R_1)\ketop_f=-0.11$,  
$\bra\sum_{i=1}^2\sum_{k=1}^N \Delta \Vint(\vec R_i- \vec r_k)
\ketop_f=0.79$. Also, we find that  
 $\bra \QR\ketop_f$ is sufficiently small 
to be ignored  (within the numerical accuracy of our simulations). 
Using these values, we estimate 
\begin{equation}
\Fenergy_{2}=-5.2. 
\label{nres1}
\end{equation}

Next, by exchanging the roles of particles {\bf 1} and  
{\bf 2} (see Fig. \ref{fig:colloid3} (b)),  
we can define $\Fenergy_1$ in the 
same way:  
\begin{eqnarray}
&& -\Fenergy_{1}\Delta R_1 =
\bra \QRR\ketopp_{f}+\bra \tilde{Q}_{\rm b}\ketopp_{f}  
+\bra\Delta\Vt(\vec R_2)\ketopp_{f}\nonumber \\ 
&+&\bra\sum_{i=1}^2\sum_{k=1}^N \Delta \Vint(\vec R_i- \vec r_k)\ketopp_{f}  
+O(\Delta R_1^2). 
\label{energyb4}
\end{eqnarray}
In Figs. \ref{fig:ut} and \ref{fig:u1u2},  we plot 
$\bra\Delta \Vt\ketopp_f$  and 
$\bra\sum_{i=1}^2\sum_{k=1}^N \Delta\Vint(\vec R_i- \vec r_k)
\ketopp_f$  as functions  
of $\Delta R_1$ in the case $f=0.3$. 
Also, in Fig. \ref{fig:heat0.3}, $\bra \Jr(t) \ketopp_f$ is plotted 
as a function of time in the case $\Delta R_1=0.25$. 
Then, obtaining 
$\bra \tilde{Q}_{\rm b}\ketopp_f=0.64$,  
$\bra\Delta\Vt(\vec R_1)\ketopp_f=-0.05$, 
$\bra\sum_{i=1}^2\sum_{k=1}^N \Delta \Vint(\vec R_i- \vec r_k)
\ketopp_f=0.79$, 
and ignoring the small contribution from $\bra \QRR\ketopp_f$, 
we estimate 
\begin{equation}
\Fenergy_{2}=5.5. 
\label{nres2}
\end{equation} 
The numerical results given in (\ref{nres1}) and (\ref{nres2}) 
are consistent 
with the law of action and reaction with respect to 
$F_1^{\rm thermo}$ and $F_2^{\rm thermo}$. While it is desirable to obtain
more precise results [The uncertainty on the values of 
$F_i^{\rm thermo}$ 
is due mainly to the difficulty in determining 
$\bra \tilde{Q}_{\rm b}\ketop_f$ and 
$\bra \tilde{Q}_{\rm b}\ketopp_f$ (see Fig. \ref{fig:heat0.3}).], 
it is clear from the present results that even if the law of action and
reaction is violated for $F_i^{\rm thermo}$ in the NESS we consider, 
the extent of this violation is
much less than that in the case of $F_i^{\rm mec}$ [recall that 
we found $\bra F_1^{\rm mec}\kets{f} = 7.0$ while 
$\bra F_2^{\rm mec}\kets{f} = -3.6$ (see Fig. 
\ref{fig:sum})].

\subsection{Statistical mechanical definition}

%
%

In this subsection, we investigate the interaction force between 
particles {\bf 1} and  {\bf 2} 
in a NESS from  a statistical mechanical point of view, employing 
the steady state 
distribution function, $\ps(\vec R_1,\{\vec r_k\};\vec R_2)$.  
It is known that 
$\ps(\vec R_1,\{\vec r_k\};\vec R_2)$ is given by  
\begin{equation}
\ps(\vec R_1,\{\vec r_k\};\vec R_2)
=\lim_{\tau \to \infty}
p_{\rm c}(\vec R_1,\{\vec r_k\};\vec R_2)
\overline{\exp \left(-\beta  \Sigma_\tau \right)},
\label{ssd0}
\end{equation} 
with 
\begin{equation}
\Sigma_\tau =\int_0^{\tau} dt 
\sum_{k=1}^N \vec f\cdot \dot{\vec r}_k,
\label{entpr}
\end{equation} 
where $\overline{A}$ denotes the 
statistical average of $A$ over noise histories under the 
condition that  the  initial condition   
$(\vec R_1(0),\{\vec r_k(0)\})$ is fixed as the argument 
$(\vec R_1,\{\vec r_k\})$ of the distribution function 
$\ps(\vec R_1,\{\vec r_k\};\vec R_2)$. 
The expression (\ref{ssd0}) indicates that the deviation of 
$\ps(\vec R_1,\{\vec r_k\};\vec R_2)$ from the canonical distribution, 
$p_{\rm c}(\vec R_1,\{\vec r_k\};\vec R_2)$, is represented by 
the entropy production $\beta \Sigma_\tau$.
This form of the steady state distribution is similar to that 
obtained by Zubarev and McLennan  \cite{Zuba,Mc}. 
(Regarding the derivation 
for stochastic systems, see  Subsection 4.2 of Ref. \cite{lrt}.)


Now, referring to the equilibrium case, we define an effective potential
$\Veff(\vec R_1;\vec R_2)$ by 
\begin{eqnarray}
&&\exp[-\beta(\Veff(\vec R_1;\vec R_2)+\Vt(\vec R_1))]
\nonumber \\ 
&=& \lim_{\tau \to \infty} 
\int \prod_{k=1}^N  d\vec r_k  
p_{\rm c}(\vec R_1,\{\vec r_k\};\vec R_2)
\overline{\exp\left(-\beta \Sigma_\tau\right)} .
\label{effs}
\end{eqnarray}
With this effective potential,   we   
 define the statistical mechanical force as 
\begin{equation}
\vec \Feff_2\equiv -\pder{\Veff(\vec R_1;\vec R_2)}{\vec R_2}.
\end{equation}
Then,  differentiating (\ref{effs}) with 
respect to $\vec R_2$, assuming that the relation 
\begin{equation}
\pder{}{\vec R_2} \overline{\exp(-\beta \Sigma_{\tau})}
=-\beta \left(\frac{\partial}{\partial \vec R_2} 
\overline{\Sigma_\tau} \right)  \overline{\exp(-\beta \Sigma_{\tau})} 
\label{assump}
\end{equation}
holds, we derive  
\begin{eqnarray}
\bra \Feff_2\kets{f}&=& \bra \Ftot_2\kets{f} \nonumber \\
&-&  \lim_{\tau \to \infty} 
\int d\vec R_1 \prod_{k=1}^N d\vec r_k\
\ps(\vec R_1,\{\vec r_k\};\vec R_2) 
\frac{\partial}{\partial \vec R_2} 
\overline{\Sigma_\tau}. 
\label{effs2}
\end{eqnarray}
Note that (\ref{assump})  is valid to  linear order in $\vec f$ 
because the both hand sides of (\ref{assump}) become
\begin{equation}
-\beta \frac{\partial}{\partial \vec R_2} 
\overline{\Sigma_\tau} +O(|\vec f|^2).
\end{equation}
We expect that (\ref{assump}) might be  approximately valid even for large 
$\vec f$, due to many-body effects.  
In the argument below, we assume (\ref{assump}).

On the other hand, using the energy balance equation obtained from the 
Langevin equations (\ref{model1}) and (\ref{model2}), we can rewrite
(\ref{energyb3}) as 
\begin{eqnarray}
\Fenergy_2&\Delta& R_2 = \bra\Ftot_2\kets{f}\Delta R_2
\nonumber \\ 
&-& \lim_{\tau \to\infty} \vec f\cdot\left(
\bra \int_0^{\tau}dt \sum_{k=1}^N \dot{\vec r}_k\ketop_f- 
\bra \int_0^{\tau}
dt \sum_{k=1}^N \dot{\vec r}_k\kets{f}\right).
\label{effs4}
\end{eqnarray}
Then, recalling that $\bra \ \ketop_f$ is the statistical average with  
respect to 
the instantaneous shift $\vec R_2\to \vec R_2+\Delta \vec R_2$ 
at $t=0$, we can write 
\begin{eqnarray}
&&\bra \int_0^{\tau}dt \sum_{k=1}^N \dot{\vec r}_k\ketop_f- 
\bra \int_0^{\tau}
dt \sum_{k=1}^N \dot{\vec r}_k\kets{f} \nonumber \\ 
&=&  \int d\vec R_1 \prod_{k=1}^N d\vec r_k\
\ps(\vec R_1,\{\vec r_k\};\vec R_2) \pder{}{\vec R_2} 
\overline{ \int_0^{\tau}dt \sum_{k=1}^N \dot{\vec r}_k}  
\Delta \vec R_2\nonumber \\ 
&+&O(|\Delta \vec R_2|^2).
\end{eqnarray} 
Substituting  this result into (\ref{effs4}) and comparing the obtained
expression with (\ref{effs2}),  we find 
\begin{equation}
\Fenergy_2=\bra\Feff_2\kets{f}. 
\label{zugoal}
\end{equation}

Thus, the equality $\Fenergy_1=-\Fenergy_2$ observed numerically  
in the last subsection suggests that  $\vec \Feff_1$ defined by 
\begin{equation}
\vec \Feff_1\equiv -\pder{\Veff(\vec R_2;\vec R_1)}{\vec R_1} 
\end{equation}
satisfies the relation 
\begin{equation}
\vec \Feff_1 = - \vec \Feff_2.  
\label{2assum1}
\end{equation}
We further assume that 
\begin{equation}
\Veff(\vec R_1;\vec R_2) = \Veff(\vec R_2;\vec R_1). 
\label{2assum2}
\end{equation} 
 Then, (\ref{2assum1}) and (\ref{2assum2}) imply that 
there exists a function $\tilde{V}_{\rm eff}(\vec R_1-\vec R_2)$  
such that 
\begin{eqnarray}
\tilde{V}_{\rm eff}(\vec R_1-\vec R_2)&=&\Veff(\vec R_1;\vec R_2)
\nonumber \\
&=& \Veff(\vec R_2;\vec R_1).
\label{veffness}
\end{eqnarray}
It is a future problem to develop a theoretical argument for 
the suggestion (\ref{2assum1}).


%
%
%

\section{Discussion}


In this paper, we have studied three types of force,  
 $\Feff_i$, $\Ftot_i$ and $\Fenergy_i$,  
in a NESS.  In equilibrium, these forces 
satisfy the relations $\bra\Feff_i\kets{f=0} 
=\bra\Ftot_i\kets{f=0}=\Fenergy_i$.  
From the results of our numerical simulations, 
we found that in the NESS that we study,    
$\Fenergy_1=-\Fenergy_2$  but $\bra\Ftot_1\kets{f}\ne-\bra\Ftot_2\kets{f}$, 
and hence the law of action and reaction holds for forces of the 
former kind but not the latter kind. 
We also demonstrated that $\Fenergy_i$ defined with respect to 
a form of the 
first law of thermodynamics hypothesized to hold for our NESS   
is equal to $\bra\Feff_i\kets{f}$ derived from the expression 
for the steady state distribution  under the assumption represented 
by  (\ref{assump}).


The law of action and reaction holds for any force given by the 
derivative of a potential function. 
Thus, the fact that this law holds with respect to the thermodynamic 
force defined by  (\ref{energyb3}) 
implies that in the NESS we considered, there may exist a 
potential function associated with this force.
It is important to clarify the physical conditions under which the 
thermodynamic force obtained
using the method employed in this paper indeed is a potential force.


In equilibrium, a potential function is related to the  
thermodynamic free energy.  
For this reason, it is natural to conjecture that even in a 
 NESS, the potential 
force measured  experimentally   is related  to to  
thermodynamics. Such a conjecture led us  
to introduce the thermodynamic force based on  the idea of Oono and 
Paniconi 
\cite{oono}.  Here we remark that the framework proposed by Oono 
and Paniconi  is embodied in the nonequilibrium  
Langevin model \cite{hatasasa}.  
In that work,  Hatano and Sasa  derived an identity  
which yields an extended second law relating the Shannon entropy with 
the excess heat they defined.  Recently, the validity of this identity 
was confirmed experimentally for  a bead system \cite{busta}.  
In the analysis employed 
in the present paper, we conjecture that $\tilde{Q}_{\rm b}$ 
defined in 
(\ref{heat3}) has a certain connection with the excess heat 
defined in Ref. \cite{hatasasa}.  However, we do not yet understand 
this relation, though apparently  the quantity $\tilde{Q}_{\rm b}$ 
employed here 
takes a simpler form than that proposed by Hatano and Sasa.


As another important subject related to the present work,
we now consider the problem of  {\it force decomposition}. As seen 
in Subsection \ref{s:sum} (see  Fig. \ref{fig:sum}), the law of 
action and reaction does not hold for   
the total mechanical force acting on the test particle,  
$\bra \vec \Ftot_i \kets{f}$.  
If the potential force in a NESS can be defined properly on 
the basis of thermodynamic considerations, it might be possible 
to decompose $\langle \vec{F}_i^{\rm mec} \rangle_f^s$ into its potential 
and non-potential parts.  Furthermore, 
because $\vec \Ftot_i$ fluctuates in time, we would like to  decompose 
  $\vec \Ftot_i$  into a  
fluctuating dissipative force, a fluctuating interaction potential 
force and   the remaining part.


With regard to  the decomposition of fluctuating forces, 
recently, a significant result has  
been obtained for a model of a Brownian particle that is driven 
by an external driving force and  subject to  
a spatially periodic potential.  As this particle is driven in one 
direction, it experiences repeated 'collisions'  with the periodic 
potential barriers. 
 Studying a finite time average of the force exerted by the periodic 
potential under nonequilibrium conditions, we have found that a 
simple orthogonality condition provides a proper force decomposition 
of the time averaged force  
into a dissipative force and a non-dissipative part \cite{HSV}.  
This result has led us to construct a 
mathematical technique for  re-expressing a Langevin equation in a 
form in which  the response function appears explicitly \cite{HHS}.  
Using this technique, an interesting equality that connects  
energy dissipation with the violation of a fluctuation-response 
relation has been proved \cite{harasasa}. Considering these 
developments,  the problem of decomposing $\bra \vec \Ftot_i \kets{f}$ 
 can be 
regarded as a natural extension of the study presented in Ref. \cite{HSV}, 
involving the elimination of the dynamical degrees of freedom. 


To summarize the most important point of this work, we have found that 
 the law of action and reaction with respect to thermodynamic  
forces has a deep connection with  fundamental aspects of 
nonequilibrium statistical mechanics. We believe that the numerical
findings  reported in this paper will stimulate further theoretical and
experimental studies.

\ack

This work was supported by  grants 
from JSPS Research Fellowships  for Young Scientists 
(Grant No. 1711222) and  the Ministry of Education, 
Science, Sports and Culture of Japan (Grant No. 16540337).

\end{document}